\documentclass[aps,prd,nofootinbib,preprint]{revtex4-1}
\usepackage{graphicx}
\usepackage{xcolor}
\usepackage{amsmath}

\def\be{\begin{equation}}
\def\ee{\end{equation}}
\def\bea{\begin{eqnarray}}
\def\eea{\end{eqnarray}}
\def\met{\slash{\!\!\!\!E}_T}

\begin{document}

\title{Implications of the Little Higgs Dark Matter and T-odd Fermions}

\author{\vspace{1cm} Chuan-Ren Chen$^{\, a,b}$,   Ming-Che Lee$^{\, a}$ and  Ho-Chin Tsai$^{\, b,c,d}$ }

\affiliation{
\vspace*{.5cm}
  \mbox{$^a$Department of Physics, National Taiwan Normal University, Taipei 116, Taiwan}\\
\mbox{$^b$ Center for High Energy Physics, Chung-Yuan Christian University} \\
\mbox{$^c$ Department of Physics, Chung Yuan Christian University, Chung-Li 320, Taiwan and}\\
\mbox{$^d$ Institute of Physics, Academia Sinica, Taipei 115, Taiwan}\\
\vspace*{1cm}}
\begin{abstract}
We study the phenomenology of dark matter in the Littlest Higgs model with T-parity after the discovery of Higgs boson. 
We analyze the relic abundance of dark matter, focusing on the effects of coannihilaitons with T-odd fermions. After determining  the parameter space that predicts the correct relic abundance measured by WMAP and Planck collaborations, we evaluate the elastic scattering cross section between dark matter and nucleon. In comparison with experimental results,   we find that the lower mass of dark matter is constrained mildly by LUX 2013 while the future XENON experiment has potential to explore most of the parameter space for both T-odd lepton and T-odd quark coannihilation scenarios.  We also study the collider signatures of T-odd fermion pair production at the LHC. Even though the production cross sections are large, it turns out very challenging to search for these T-odd fermions directly at the collider because the visible charged leptons or jets are very soft.  Furthermore, we show that, with an extra hard jet radiated out from the initial state, the T-odd quark pair production can contribute significantly to mono-jet plus missing energy search at the LHC.  
\end{abstract}

\maketitle

\section{Introduction}
\label{sec: intro}

Recently, a new scalar with mass of about $125$ GeV has been discovered at the Large Hadron Collider (LHC)~\cite{Aad:2012tfa,Chatrchyan:2012ufa}, it is still unclear at this moment whether or not this new particle is the Higgs boson in the Standard Model (SM). Due to the large radiative corrections in Higgs boson mass parameter in the SM, a very precise cancelation must take place in order to have the Higgs boson mass in the electroweak scale if the SM is valid up to Planck scale. This so-called naturalness problem serves a driving force for theorists to propose many solutions, the leading new physics beyond the SM is supersymmetry. Another elegant idea is Little Higgs mechanism, in which the light Higgs boson is realized as a pseudo-Goldston boson. With collective symmetry breaking mechanism \cite{ArkaniHamed:2001nc} (also see~\cite{Schmaltz:2005ky, Perelstein:2005ka} for review), the global symmetry breaking scale $f$ can be at $\cal O$(1 TeV) without a fine tuning.  The one-loop quadratic divergences induced by the SM particles are exactly cancelled by new fermions, gauge bosons and scalars. An economical model is Littlest Higgs model~\cite{ArkaniHamed:2002qy}. However, in order to satisfy the electroweak precision measurements, the scale $f$ is required to be greater than $4$ TeV~\cite{Csaki:2002qg,Hewett:2002px,Chen:2003fm} and a fine tuning to the light Higgs boson mass is reintroduced.  One of the solutions is to embed a discrete symmetry, called T-parity, into the model~\cite{Cheng:2003ju,Cheng:2004yc,Low:2004xc} so that no mixing between new particles which  are assigned with T-parity odd and the SM particles which are T-parity even. The corrections to electroweak observables are therefore all loop-induced. As a result, the scale $f$ can be as low as about $500$ GeV~\cite{Hubisz:2005tx} and the LHC has great potential to examine the model.The phenomenology of the Littlest Higgs model with T-parity has been studied in the literature~\cite{Hubisz:2004ft,LHT:pheno} . Moreover, the T-parity also ensures the stability of the lightest T-odd particle (LTP) that is naturally can be the candidate of dark matter if it is charge-neutral and colorless. 
There are two possible candidates of dark matter in Littlest Higgs model with T-parity: T-odd partner of the hypercharge gauge boson $A_H$~\cite{Hubisz:2004ft,Birkedal:2006fz, Asano:2006nr} and the T-odd partner of neutrino $\nu_H$. 

The current dark matter relic abundance in our universe has been measured by WMAP \cite{Hinshaw:2012aka} and recently by Planck \cite{Ade:2013zuv} with the combined value
 \be
 \Omega_{DM}h^2 = 0.1199 \pm 0.0027.
 \label{eq:planck}
 \ee 
We found that it is possible to explain the measurement of dark matter relic abundance with $\nu_H$ dark matter. However, the direct search experiment excludes such a possibility. The reason is that the coupling strength of  $\nu_H$ to $Z$-boson is similar to that of $Z$-boson to SM fermions, therefore, the cross section of elastic scattering between $\nu_H$ and nucleus is about $4\sim5$ order of magnitude larger than the current experimental search bound. This situation has been also noticed in the case of  KK-neutrino dark matter in Universal Extra-dimensional model~\cite{Servant:2002hb}. Hence, we will focus on $A_H$ dark matter in our study. To predict the relic abundance of dark matter, we should calculate the dark matter annihilation cross section. The pair annihilation of two $A_H$ dark matter is mainly through the Higgs boson in the $s$-channel to SM particle final states. With a $125$ GeV Higgs boson, the annihilation cross section of a pair of $A_H$ is too small to fit the measurement of dark matter relic density (Eq.~(\ref{eq:planck})) unless the mass of $A_H$ is very close to half of the Higgs boson mass  in order to have resonance enhancement. Such restriction can be lifted when various coannihilation channels are included. In this paper, we take coannihilation processes into account  and explore the effects in direct search experiments and implications of LHC phenomenology.

The rest of paper is organized as follows. We begin in Sec.~\ref{sec: model} with a brief introduction of the Littlest Higgs model with T-parity.   In Sec.~\ref{sec: dm}, we study the coannihilation with light T-odd fermions, separately for T-odd leptons and for T-odd quarks. We first determine the parameter space that can fit the current measurement of dark matter relic abundance, and then calculate the dark matter elastic scattering with nucleus.  The LHC study is present in Sec.~\ref{sec: lhc}. We calculate the production cross section of T-odd fermion. We point out that the T-odd fermion production contribute significantly to the mono-jet with missing energy search for dark matter at the LHC.  Our conclusion appears in Sec.~\ref{sec: con}.

\section{Littlest Higgs Model with T-parity}
\label{sec: model}
The Littlest Higgs model is a SU(5)/SO(5) nonlinear sigma model where Higgs boson is a pseudo Nambu-Goldstone boson. The global symmetry SU(5) is broken down to SO(5) by a $5\times5$ symmetric tensor $\Sigma_0$ at the scale $f$, where 
\be
\Sigma_0 = \begin{bmatrix}
0&0&0&1&0\\
0&0&0&0&1\\
0&0&1&0&0\\
1&0&0&0&0\\
0&1&0&0&0\\
\end{bmatrix}.
\ee
 A subgroup of $SU(5)$, $[SU(2)\times U(1)]_1\times[SU(2)\times U(1)]_2$, is gauged, and is broken down to the diagonal $SU(2)\times U(1)$ that is identified as SM $SU(2)_L\times U(1)_Y$.  
A discrete symmetry, $T$-parity, that governs the transformation of fields between $[SU(2)\times U(1)]_1$ and $[SU(2)\times U(1)]_2$  is introduced. 
 Two different linear combinations of fields under the gauged $[SU(2)\times U(1)]_1$ and $[SU(2)\times U(1)]_2$ define the SM particles and extra heavy particles. For gauge bosons sector, the heavy gauge bosons take the form $W^{a}_{H\mu}=(W^a_{1\mu} -W^a_{2\mu})/\sqrt{2}$ and $A_{H\mu}=(B_{1\mu}-B_{2\mu})/\sqrt{2}$ and are odd under T-parity transformation, where $W_{1,2\mu}^a$ and $B_{1,2\mu}$ are gauge boson fields of $SU(2)_{1,2}$ and $U(1)_{1,2}$, respectively, and $a=1,2,3$. For fermion sector, the T-odd fermion is assigned as $\psi_H = (\psi_1 + \psi_2)/\sqrt{2}$, while the SM fermion is $\psi_{SM}=(\psi_1 - \psi_2)/\sqrt{2}$ that is even under T-parity transformation, where $\psi_{1,2}$ are doublet under $SU(2)_{1,2}$. As a result, there is no mixing between heavy T-odd particles and SM particles and the all the tree-level contributions to electroweak precision tests are forbidden. 
 The stringent constraints on scale $f$  can be relaxed and the model is then testable at the LHC.  
 
To cancel the one-loop quadratic divergence indued by SM top quark in Higgs boson mass correction, another set of singlet $U_1$ and $U_2$ is introduced in such a way that a new T-even particle $T_+=(U_1-U_2)/\sqrt{2}$ exactly cancels the contribution of SM top quark loop. And the other combination $T_-=(U_1+U_2)/\sqrt{2}$ is odd under T-parity.   
The masses of these new gauge bosons and fermions, including the corrections after SM electroweak symmetry is broken, are 
 \bea
 M_{A_H} = \frac{g^\prime f}{\sqrt{5}}\left[ 1- \frac{5v^2}{8 f^2}\right],~~
 M_{W_H} = gf\left[1-\frac{v^2}{f^2}\right],\\\nonumber
 M_{d_H} = \sqrt{2}\kappa_q f,~~
 M_{u_H} = \sqrt{2}\kappa_q f \left[ 1-\frac{v^2}{8f^2}\right],\\\nonumber
 M_{\ell_H} = \sqrt{2}\kappa_\ell f,~~
 M_{\nu_H} = \sqrt{2}\kappa_\ell f \left[ 1-\frac{v^2}{8f^2}\right],\\\nonumber
 M_{T_+} = \sqrt{\lambda_1^2 + \lambda_2^2} f\left[1-\frac{\lambda_1^2 \lambda_2^2}{2(\lambda_1^2 +\lambda_2^2)^2}\frac{v^2}{f^2} \right],~~
 M_{T_-} = \lambda_2 f, 
 \eea
where $g$ and $g^\prime$ are  respectively the gauge couplings for SM $SU(2)_L$ and $U(1)_Y$,  $\kappa_q$, $\kappa_\ell$, $\lambda_1$ and $\lambda_2$\footnote{$\lambda_1$ and $\lambda_2$ can be related by the mass of top quark.} are free parameters in the Lagrangian that generates masses for heavy fermions. Here, $u_H$ ($d_H$) denote the T-odd partners of the SM up-(down-)type quarks, $\nu_H$ are T-odd partners of the neutrinos and $\ell_H$ are T-odd partners of the charged leptons. Note that the masses of heavy gauge bosons depend on $f$ only, and T-odd fermions rely on an additional parameter $\kappa_q$ or $\kappa_\ell$. For simplicity, we assume the universal $\kappa_\ell$ ($\kappa_q$) for T-odd partners of three generations of leptons (quarks) of SM unless otherwise specified. We can see that, when  $\kappa_q$ is smaller than $0.11$, the up-type T-odd quarks become the lightest T-odd particle and are stable, which conflicts  with results of dark matter searches.  For $\kappa_\ell \lesssim 0.11$, T-odd $\nu_H$ can be the dark matter candidate. However, such a possibility is excluded by the direct searches as we mentioned previously. Therefore, in the following studies, we take both $\kappa_q$ and $\kappa_\ell$ to be larger than $0.11$. We should refer readers who are interested in details of the Littlest Higgs model with T-parity to Ref.~\cite{Cheng:2003ju,Cheng:2004yc,Low:2004xc}.

\section{Relic abundance and direct detection}
\label{sec: dm}

In this section, we study the dark matter $A_H$ in Littlest Higgs Model with T-parity (LHT) in comparison with current dark matter experiments. We will first calculate the relic abundance and then the direct detections.  All the calculations shown here are performed with MicrOMEGAs3.1 package~\cite{Belanger:2013oya}. 

The relic density of dark matter today is determined by its annihilation processes in the non-relativistic limit. In LHT, a pair of $A_H$ mainly annihilate through a Higgs boson as the mediator in $s$-channel  to a pair of $b$ quarks, $W/Z$ bosons, Higgs bosons or top quarks, depending on the mass of $A_H$~\cite{Hubisz:2004ft,Birkedal:2006fz,Asano:2006nr}. When $A_H$ is lighter than $W$-boson, it annihilates to $b$ quarks, however, when $W$-boson channel is open, a pair of $W$-boson final state is always dominant. With Higgs boson mass $m_H=125~\rm{GeV}$, $A_H$ has to be heavier than $m_H/2=62.5~\rm{GeV}$ so that the decay of Higgs boson into a pair of $A_H$ is kinematically forbidden. Otherwise, if the channel $H\to A_H A_H$ is opened, it always dominates the Higgs boson decay branching ratios and conflicts with the current limit of invisible decay branching ratio of Higgs boson~\cite{invisible,Aad:2014iia}. In our study, we vary $f$ between $\sim 480$ GeV and $\sim 2$ TeV which corresponds to $62.7~\rm{GeV}\lesssim m_{A_H}\lesssim312.4~\rm{GeV}  $.

\begin{figure}[htbp]
\begin{center}
\includegraphics[scale=0.85,clip]{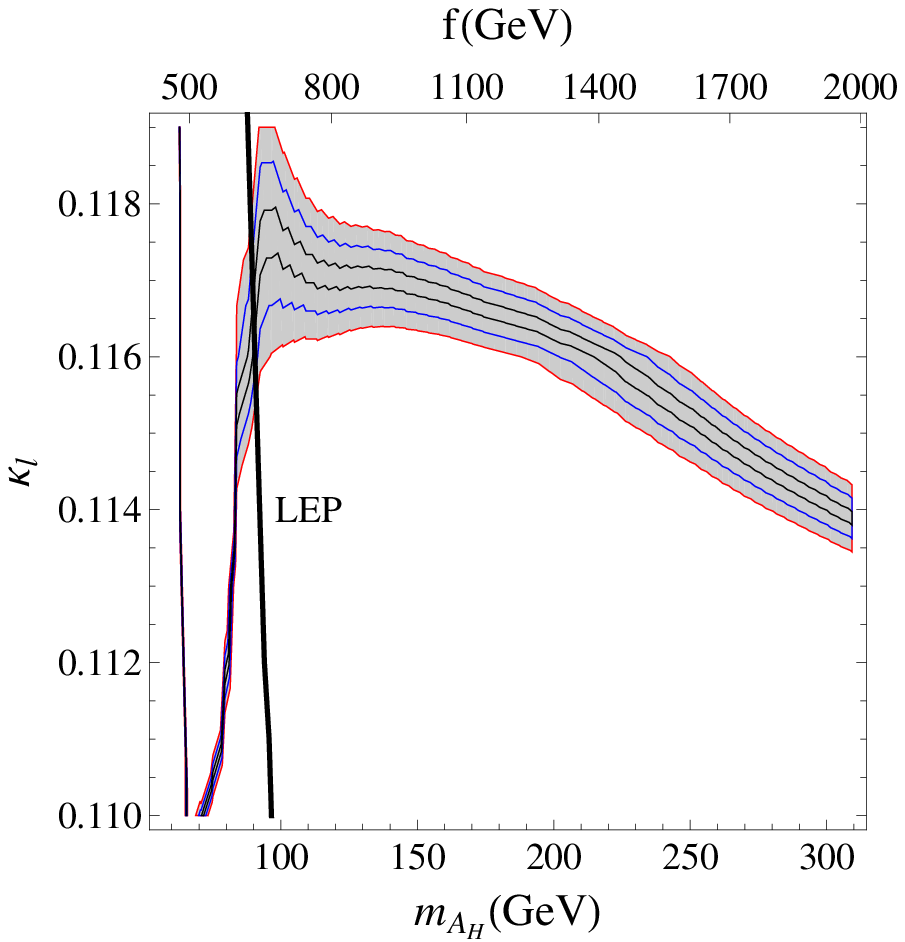}
\includegraphics[scale=0.85,clip]{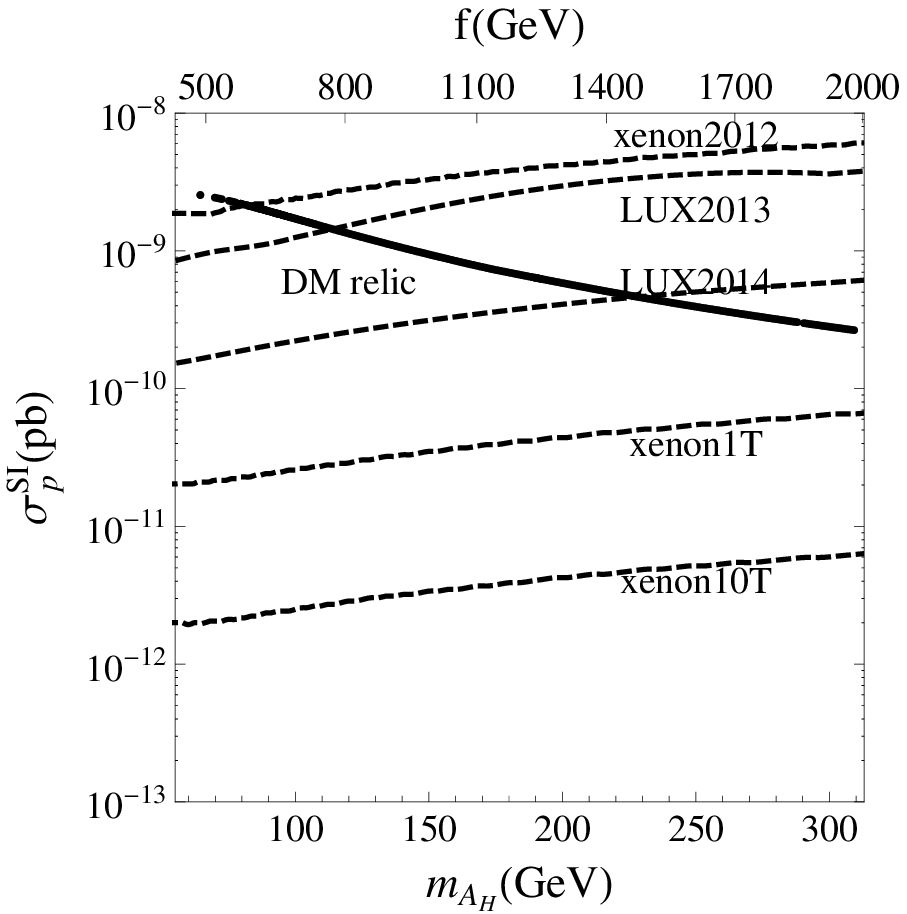}
\caption{Left: The parameter space of ($\kappa_\ell$, $m_{A_H}$ ) that predicts dark matter relic abundance that is consistent with observation Eq.~(\ref{eq:planck}) within $1\sigma, 3\sigma, 5\sigma$, shown respectively in black, blue and red lines. The thick black line shows lower limit of $f$ from searches of heavy charged particles at LEP II. Right: The one-nucleon-normalized spin-independent elastic cross-section of heavy photon $A_H$ scattering  off the nucleon together with the current experimental limits of XENON100 (2012)~\cite{Aprile:2012nq}, LUX(2013)~\cite{Akerib:2013tjd}. The projective limits of LUX(2014), XENON 1T\cite{Aprile:2012zx}, and XENON10T~\cite{xenon10T} are also shown.}
\label{fig:relic_l}
\end{center}
\end{figure}

In the case that $A_H$ is much lighter than other new heavy particles, in order to have right pair annihilation cross section to fit the current dark matter relic abundance measurement,  c.f. Eq.(\ref{eq:planck}),
the mass of $A_H$ is required to be just slightly heavier than half of Higgs boson mass $m_{A_H}\gtrsim m_H/2=62.5~\rm{GeV}$.
However, the $A_H$-nucleon elastic scattering cross section is about $10^{-9}~\rm{pb}$ that is in tension with current result form LUX~\cite{Akerib:2013tjd} and will be certainly examined by projected LUX 2014 data.
When $A_H$ becomes heavier, $A_H$ pair annihilation cross section drops quickly and the corresponding relic abundance  will be too large to agree with the observation. One possible solution to enlarge the dark matter annihilation cross section is to include coannihilation effects with T-odd fermions~\cite{Wang:2013yba}. For coannihilation to take place, T-odd fermions should be as light as dark matter $A_H$. We will demonstrate the coannihilation of T-odd leptons and T-odd quarks separately.

When T-odd leptons participate in coannihilation processes,  one more parameter $\kappa_\ell$ needs to be considered in addition to $f$. We show in left panel of Fig.~\ref{fig:relic_l} the parameter space of $\kappa_\ell$ and $f$ that predicts the right  dark matter relic abundance. The black, blue and red curves denote that the relic abundance is consistent with observation within $1\sigma$, $3\sigma$, and $5\sigma$ level.   The value of $\kappa_\ell$ is bounded from below by $0.11$ below which the $A_H$ is no longer the LTP and can not serve as the dark matter candidate. In calculations, we set all of the T-odd quarks to be much heavier than $A_H$  and are irrelevant in annihilation cross section. 
 The nearly vertical narrow band in the left part of the plot corresponds to the case that $m_{A_H}$ is just slightly heavier than $m_H/2$ where the resonance effect is significant. However, due to narrow width of Higgs boson, when $A_H$ becomes heavier, $A_H$ pair annihilation cross section drops quickly and  significant contributions of coannihilation processes involving light T-odd leptons are needed to enlarge the total annihilation cross section of dark matter. 
  When $A_H$ is heavier than $W$-boson, pair annihilation cross section becomes larger since  $A_H A_H \to W^+ W^-$ is opened. Therefore, the contribution from T-odd lepton coannihilation becomes less important than the previous case. As a result,  we see that   $\kappa_\ell$ goes higher, meaning the mass gap between the T-odd lepton and $A_H$ is larger so that the coannihilation becomes less efficient.  
Furthermore, the nearly vertical  black  line situated at $m_{A_H} \simeq 92~\rm{GeV}$ (or $f\!\sim 650~\rm{GeV}$) is the lower mass limit of T-odd lepton set by the heavy charged lepton searches at LEP, below which the T-odd pair production is too large and contradicts the null result\footnote{The LHT limit is obtained by scaling the results from LEP2 SUSY Working Group on combined LEP Selectron/ Smuon/ Stau Results, 183 - 208 GeV with ALEPH, DELPHI, L3, OPAL Experiments \cite{lep2}.}.

In the right panel of Fig.\ref{fig:relic_l} we show the spin-independence scattering cross-section of heavy photon $A_H$ with nucleon using the parameter space of ($\kappa_\ell$, $f$) which is compatible to the correct relic. 
The scattering process  is dominated by Higgs t-channel mediated while the contributions from heavy T-odd quarks is small
due to the small couplings of $A_H$ to T-odd quarks and the heaviness of T-odd quarks~\cite{Birkedal:2006fz}. 
We also show the experimental limits of XENON in 2012~\cite{Aprile:2012nq}, LUX 2013~\cite{Akerib:2013tjd}, LUX expected in 2014, projective XENON 1T~\cite{Aprile:2012zx}, and projective XENON 10T~\cite{xenon10T} for comparison. As we see that the predicted SI cross-section monotonically  decreases from $\sim 2\times10^{-9}~\rm{pb}$ for $m_{A_H}\gtrsim m_H/2$ down to $2\times 10^{-10}~\rm{pb}$ for  $m_{A_H}\simeq 310~\rm{GeV}$. It is clear that the LUX 2013 result disfavors the region where $A_H$ is lighter than $\sim 120~\rm{GeV}$ which is stronger than the LEP bound shown in left panel. In the near future, for example, projective LUX 2014 can explore the mass of $A_H$ up to about $230~\rm{GeV}$ and  XENON 1T can test all the model parameter space we are interested in.
 
For the scenario of coannihilation with T-odd quarks, we notice   that if T-odd quarks are degenerate, the decay width of T-odd partner of top quark $t_H$ is so narrow that it will leave a displaced vertex inside the detector, which contradicts with current data~\cite{Chatrchyan:2013oca}. The reason is that the 2-body $t_H\to A_H t$ and 3-body $t_H \to A_H b W^+$ decay modes   are kinematically forbidden since the mass difference between $t_H$ and $A_H$ is much smaller than top quark mass and $W$-boson mass. Therefore the only decay channel for $t_H$ is $t_H\to A_H b f\bar{f}^\prime$, where $f$ and $f^\prime$ are SM light fermions. Numerically, the decay width of $t_H$ is at the order of $10^{-14}$ GeV in the coannihiliation region. Hence,  we focus on the situation that the T-odd partners of third generation quarks are much heavier than the first two generations and are irrelevant in coannihilation processes. Also, we set all of the T-odd leptons to be much heavier than dark matter $A_H$.

\begin{figure}[htbp]
\begin{center}
\includegraphics[scale=0.85,clip]{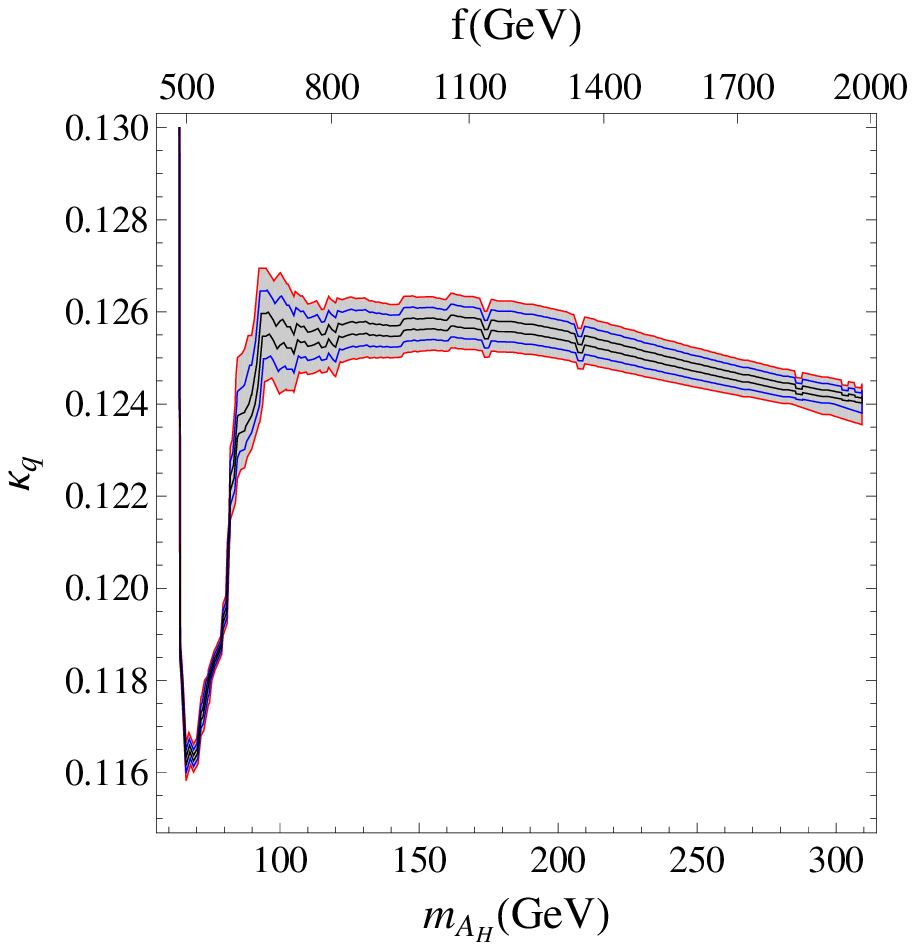}
\includegraphics[scale=0.85,clip]{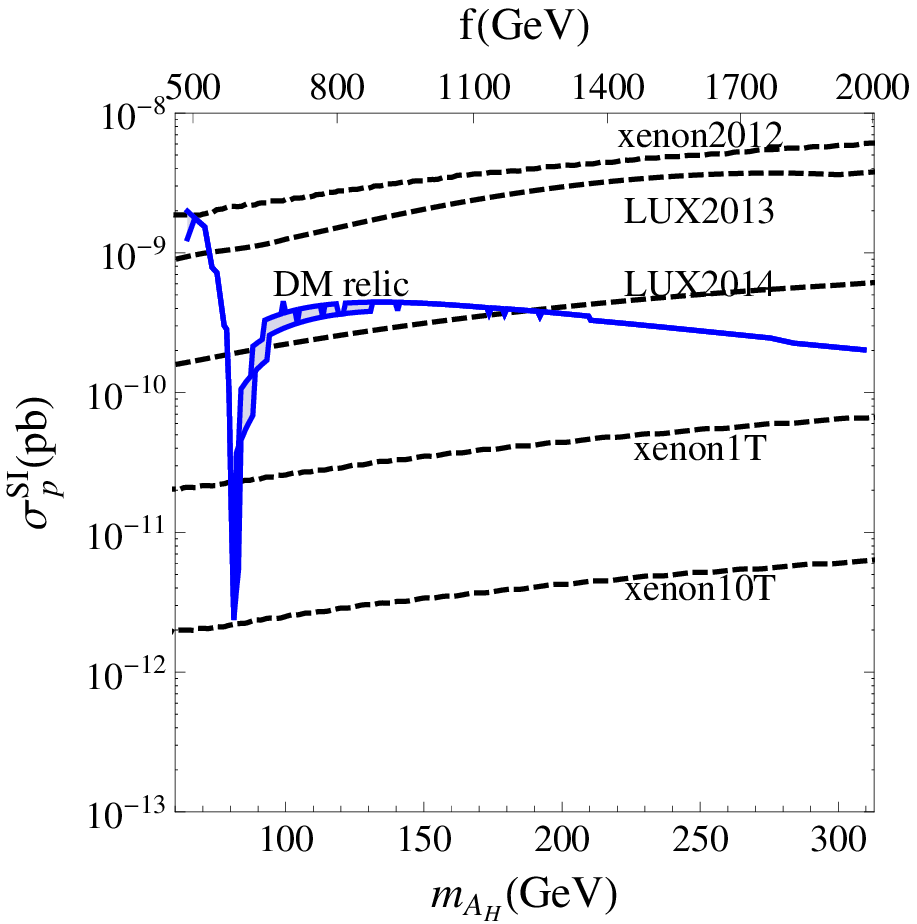}
\caption{Left: The parameter space of ($\kappa_q$, $m_{A_H}$ ) that predicts dark matter relic abundance that is consistent with observation Eq.~(\ref{eq:planck}) within $1\sigma, 3\sigma, 5\sigma$, shown respectively in black, blue and red lines. Right: The one-nucleon-normalized spin-independent elastic cross-section of heavy photon $A_H$ scattering  off the nucleon together with the current experimental limits of XENON100 (2012)~\cite{Aprile:2012nq}, LUX(2013)~\cite{Akerib:2013tjd}. The projective limits of LUX(2014), XENON 1T\cite{Aprile:2012zx}, and XENON10T~\cite{xenon10T} are also shown.}
\label{fig:relic_q}
\end{center}
\end{figure}

The left panel of Fig.~\ref{fig:relic_q} shows the parameter space of $(\kappa_q, f)$ that agrees with the dark mater relic abundance. The region between black (blue, red) lines is consistent with measurement within $1(3, 5)~\sigma$ level. Similar to the case of T-odd lepton coannihilation, as we explained previously, the sharp dropoff  at $m_{A_H}\gtrsim 62.5$ GeV is due to the fact that the cross  section of $A_H$ pair annihilation is dropping very quickly when $A_H$ is away the resonance of Higgs boson. Therefore, light T-odd quarks are needed to join the coannihilation to compensate. The rising is because the $W$-boson final state is opened in $A_H$ pair annihilation and enlarges the annihilation cross  section.  Then, the mass gap between T-odd quarks and $A_H$ should be larger to suppress the contributions of coannihilaiton processes.

Shown in the right plot of Fig.\ref{fig:relic_q} is the predicted spin-independent cross-section of $A_H$ scattering off nucleon, using the parameter space corresponding to the correct relic abundance of dark matter in the left panel of Fig.~\ref{fig:relic_q}. In addition to the Higgs-boson-exchanged $t$-channel diagrams in $A_H$-nucleon scattering, as we mentioned in previous T-odd lepton coannihilation case, the diagrams which involve T-odd quarks also play an important role since the T-odd quarks now can be as light as about $100~\rm{GeV}$~\cite{Birkedal:2006fz}. Therefore, the effects of T-odd quarks can be as significant as the one involving Higgs boson. The amplitudes between diagrams with T-odd quark exchanged interference destructively between $s$-channel and $t$-channel and may become negative in some portions of parameter space. The sharp drop-off structure at $m_{A_H}\sim 80~ \rm{GeV}$ is due to the fact that the total amplitude of the T-odd quark diagrams is negative and  destructs  the amplitude of the Higgs boson diagram. 

In comparison with the experimental limits, LUX 2013 excludes $m_{A_H} \lesssim 70~\rm{GeV}$, and the expected LUX data in 2014 has the sensitivity up to $m_{A_H} \simeq 200~\rm{GeV}$. In the future, the projective XENON 1T and XENON 10T can explore whole parameter except for the region where $m_{A_H}\simeq m_W$.

\section{Implications at the LHC}
\label{sec: lhc}

\begin{figure}[htbp]
\begin{center}
\includegraphics[scale=0.4,clip]{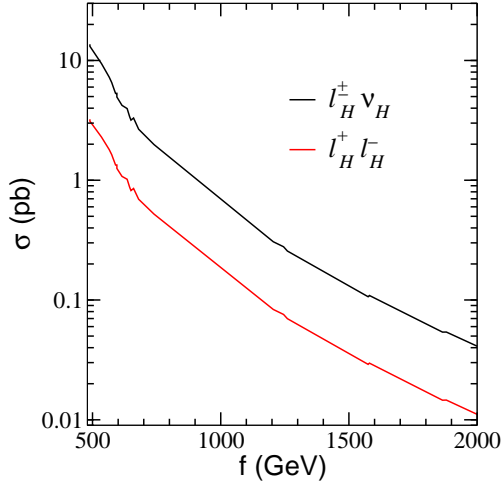}
\caption{Total production cross section of T-odd lepton pairs at the $8$ TeV LHC.  }
\label{fig:xs_TL}
\end{center}
\end{figure}

\begin{figure}[htbp]
\begin{center}
\includegraphics[scale=0.3,clip]{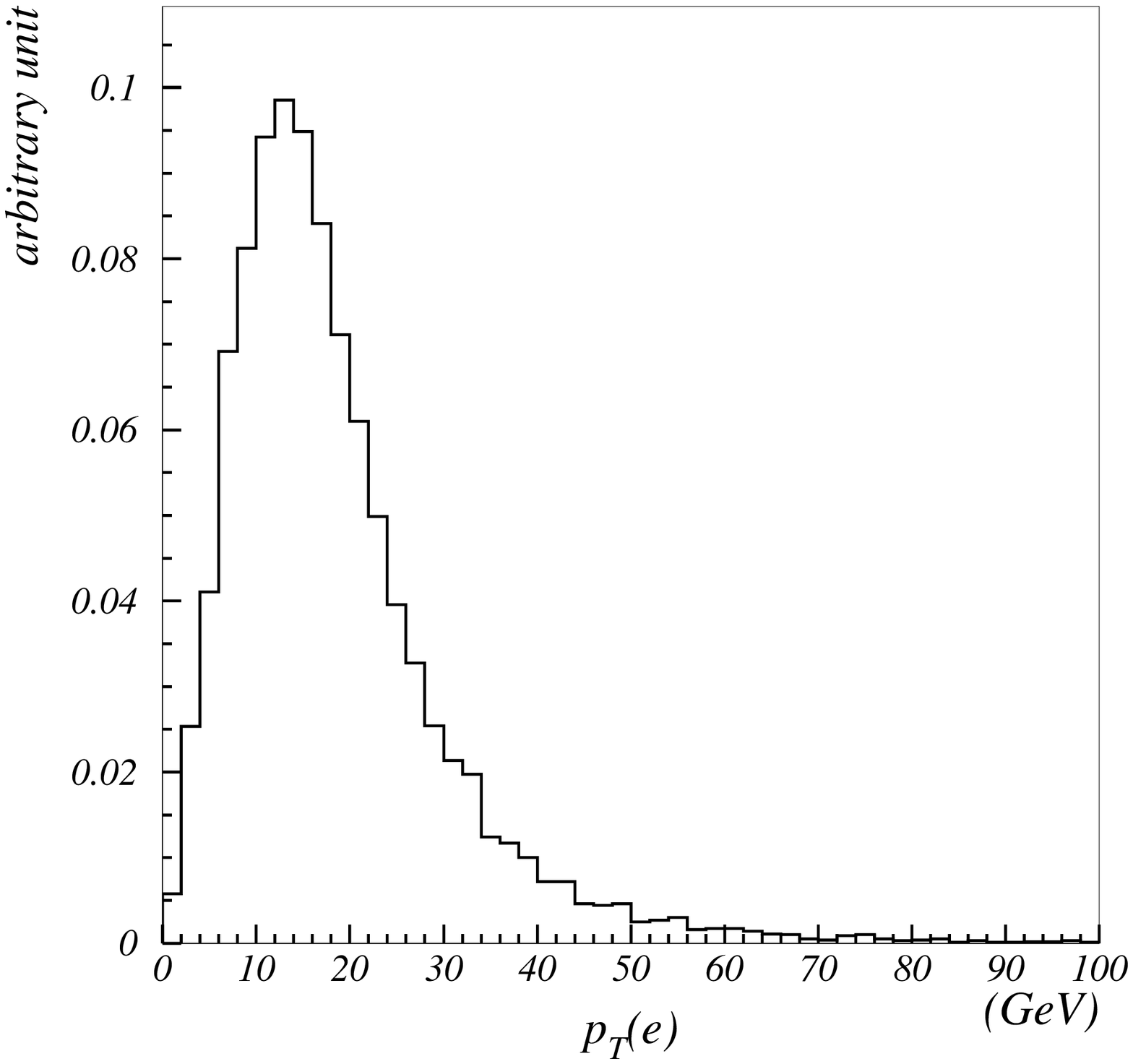}
\includegraphics[scale=0.3,clip]{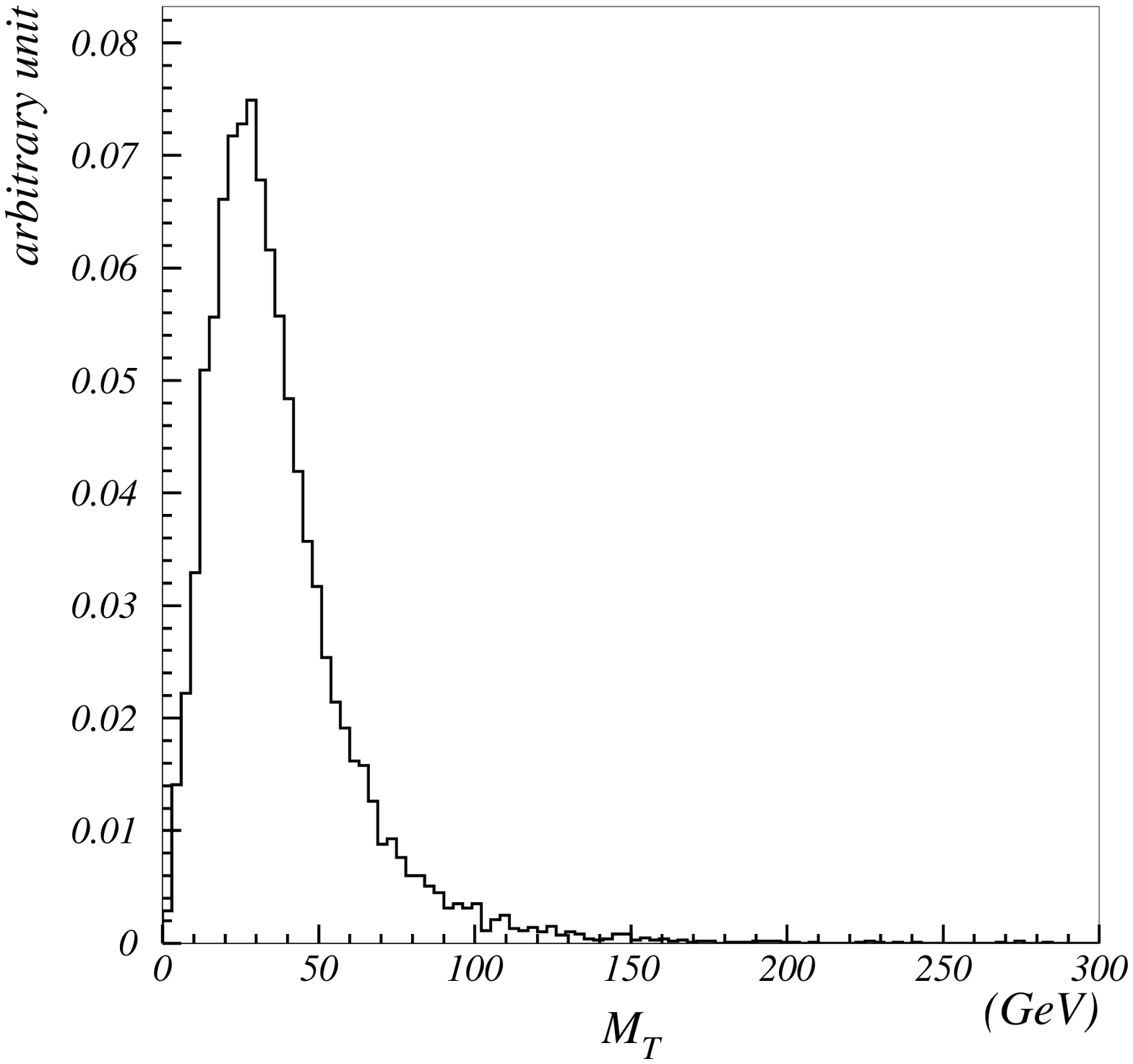}
\caption{Left: Normalized distribution of transverse momentum of $\ell^\pm$ in $pp\to \ell^\pm_H(\to \ell^\pm A_H) \nu_H (\to \nu A_H)$ production at the LHC. Right:  Normalized distribution of transverse mass in $pp\to \ell^\pm_H(\to \ell^\pm A_H) \nu_H (\to \nu A_H)$ production at the LHC.}
\label{fig:kin_TL}
\end{center}
\end{figure}

In the co-annihilations region,  The T-odd fermion $f_H$ becomes the second lightest T-odd particle and decays $100 \% $ to a dark matter $A_H$ and its SM partner fermion $f$, among which $A_H$ contributes to missing transverse momentum (MET) in the collider signatures. For example, a T-odd electron $e_H^-$ decays into $A_H$ and an electron $e^-$. Since the T-odd fermions are light, LHC can copiously produce them. First, we study the case of light T-odd leptons. As shown in Figure \ref{fig:xs_TL} , we use CalcHEP~\cite{Belyaev:2012qa} package with CTEQ6L~\cite{Pumplin:2002vw} parton distribution functions (PDF) to calculate the total production cross sections of T-odd lepton pairs at 8 TeV LHC. The cross section of $\ell_H^\pm \nu_H$ can be as large as $13~pb$ when $f$ is about $500$ GeV, where $\ell = e$, or $\mu$ or $\tau$. The collider signature of $\ell_H^\pm \nu_H$ is single charged lepton plus MET $\met$, which is the same as generic $W'$ search in leptonic decay modes. The current limit for $W'$ in single charged lepton plus MET final state requires that the $W'$  should be heavier than about $2.9$ TeV when $W'$ with SM couplings to SM fermions~\cite{Chatrchyan:2013lga}, which, in principle, can be used to interpret the constraints on T-odd lepton mass. However, since the mass gap between T-odd lepton $\ell_H^\pm$ and dark matter $A_H$ is small, the charge lepton $\ell^\pm$ in the decay $\ell_H^\pm \to \ell^\pm A_H$ is soft. As shown in right plot of Fig.~\ref{fig:kin_TL}, the transverse momentum of $\ell^\pm$ peaks around $10$ GeV. After imposing selection cuts in $W'$ search~\cite{Chatrchyan:2013lga}, the  signal of $p p \to \ell_H^\pm \nu_H\to\ell^\pm\nu A_H A_H$ is entirely cut off, especially the high transverse momentum  cut on charged lepton $p_T^\ell > 40~\rm{GeV}$ and hard transverse mass cut $M_T = \sqrt{2 p_T^\ell \cdot \met\cdot(1-\cos\Delta\phi_{\ell \nu})}>1~\rm{TeV}$. For $\ell_H^+\ell_H^-$ pair production at the LHC, the signal is $\ell^+\ell^- A_H A_H$ after T-odd leptons decay. Such dilepton plus MET signal has been searched at the LHC for slepton or chargino pair production in supersymmetry~\cite{atlas:dileptonMET}. However, due to high transverse momentum cuts on charged leptons $p_T^{\ell_{1 (2)}} > 35 (20)~\rm{GeV}$ and high $m_{T2}$ cut ($m_{T2}>90~\rm{GeV}$)~\cite{atlas:dileptonMET}, most of the signal of $\ell_H^+\ell_H^-$ do not pass the event selection. Therefore, the current searches  for heavy colorless charged particles in single charge lepton plus MET or in dilepton plus MET have no sensitivity to light T-odd lepton pair production in coannihilation region. It is quite challenging to directly search for them because of the soft charged lepton in the final state.

\begin{figure}[tbp]
\begin{center}
\includegraphics[scale=0.42,clip]{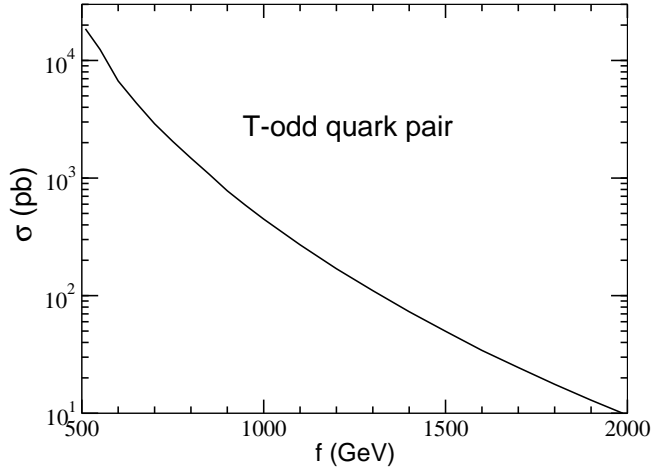}
\caption{The production cross section of a pair of T-odd quarks $q_H\bar{q}_H$ at the LHC with $8$ TeV center-of-mass energy. The cross section is the sum of the T-odd partners of the first two generation quarks. }
\label{fig:collider_q1}
\end{center}
\end{figure}

\begin{figure}[htbp]
\begin{center}
\includegraphics[scale=0.4,clip]{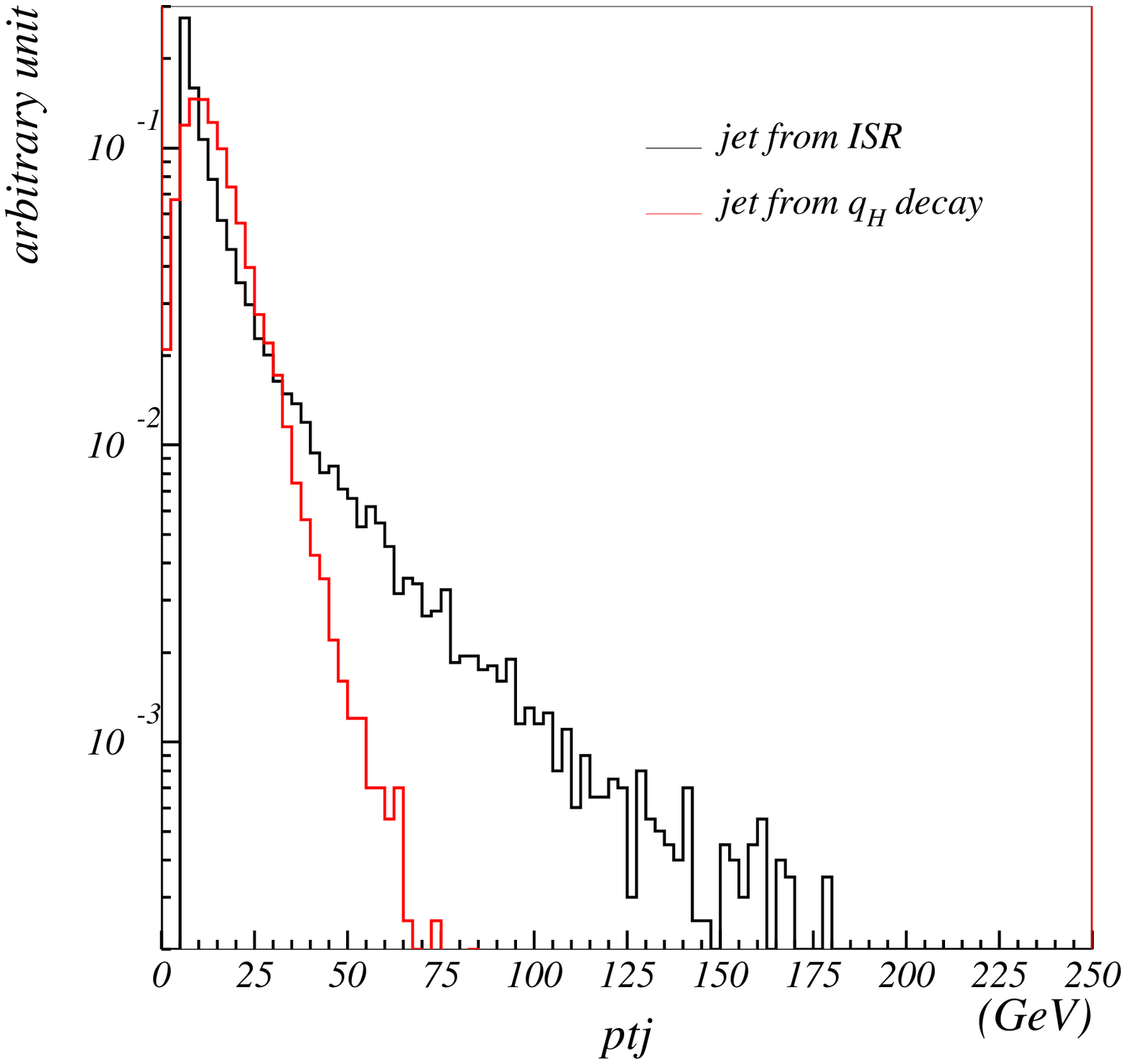}
\includegraphics[scale=0.4,clip]{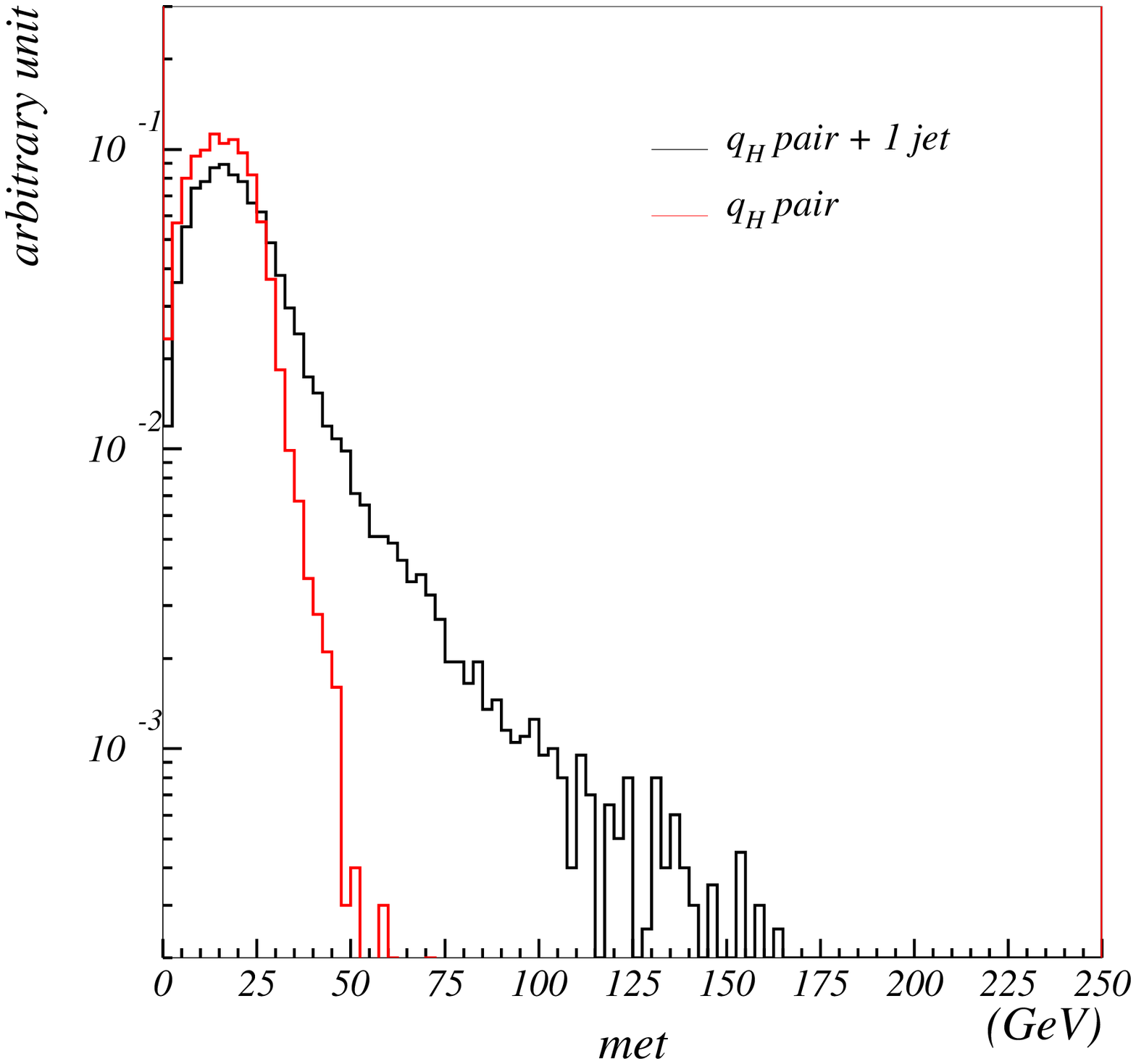}
\caption{Left: Normalized distributions of transverse momentum of the jet from initial state radiation in $pp\to q_H\bar{q}_H j$ (black histogram) and the jet from T-odd quark decay in $pp\to q_H \bar{q}_H$ (red histogram);  Right: Normalized distributions of missing transverse energy  in $pp\to q_H\bar{q}_H j$ (black histogram) and  in $pp\to q_H \bar{q}_H$ (red histogram). }
\label{fig:collider_q2}
\end{center}
\end{figure}

\begin{figure}[tbp]
\begin{center}
\includegraphics[scale=0.4,clip]{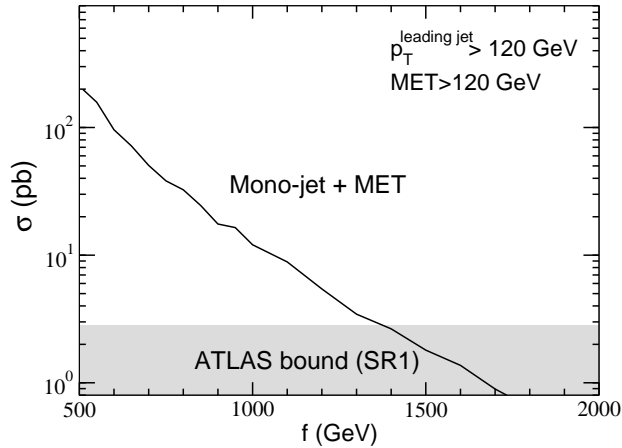}
\caption{The predicted cross section of mono-jet plus MET events from process of $pp\to q_H \bar{q}_H j$ at the LHC, using  the event selection criteria in signal region SR1. The grey shaded area is excluded by the ATLAS\cite{ATLAS:2012zim}. }
\label{fig:collider_q3}
\end{center}
\end{figure}

We now turn to study the coannihilation region where T-odd quark is light. The main production mechanism of T-odd quarks is though QCD interaction. After summing over first two generations, the total cross section of pair production of T-odd quarks  $p p \to q_H \bar{q}_H$ at the LHC with $8$ TeV center-of-mass energy is shown in Fig.~\ref{fig:collider_q1}. The cross section can be as large as about $2\times 10^4$ pb when $f\simeq 500$ GeV and decreases down to $\sim 10$ pb when $f$ is $2$ TeV.  The collider signature of T-odd quark pairs at the LHC is $2$ jets plus MET, which can mimic the signal of  squark or gluino pairs in supersymmetry. One may expect that the current data of dijet plus MET will testify this parameter space of LHT. However, similar to the charged lepton in T-odd lepton production, the jet in T-odd quark production at the LHC is very soft and most of the signals won't pass the jet ($p_T^{leading} > 130~\rm{GeV}$) and MET ($\met > 160~\rm{GeV}$) selection criteria  in dijet plus MET search~\cite{atlas:jetsMET}. As a result, the current dijet plus MET search for new physics at the LHC has no constraint on light T-odd quark scenario in the region of coannihilation.  Instead, we consider the situation that a hard jet radiated from the initial state $pp\to q_H\bar{q}_H j$. As we see in Fig.~\ref{fig:collider_q2}, the transverse momentum distribution of jet radiated from the initial state decreases slower than the one from the decay of T-odd quark. Furthermore, the MET also shifts to higher value when there  exists an extra hard  jet. Therefore, the process of a T-odd quark pair plus one jet can contribute to collider search for new physics, like dark matter, in mono-jet plus MET channel. We calculate the T-odd pair with an extra jet from initial state using Madgraph 4~\cite{Alwall:2007st} and compare with 
results of  monojet plus missing transverse momentum final states at the  LHC by ATLAS collaboration~\cite{ATLAS:2012zim}. The event selection criteria in the analysis are summarized as follows:
\begin{itemize}
\item MET $\met > 120$ GeV.
\item Leading jet with transverse momentum $p_T > 120 ~\rm{GeV}$ and rapidity $|\eta|<2.0$.
\item At most two jets with transverse momentum $p_T > 30$ GeV and rapidity $|\eta| < 4.5$.
\item If there are more than one jet, the azimuthal angle difference between second leading jet and MET $\Delta\phi(j^{\rm 2nd}, \met)>0.5$.
\item Lepton veto: no isolated electron (muon) with $p_T > 20$ GeV ($p_T>7$ GeV) and rapidity $|\eta|<2.47$ ($|\eta|<2.5$).
\end{itemize}
Furthermore, four signal regions denoted as SR1, SR2, SR3, SR4 are defined with different cuts on $p_T>120~\rm{GeV},~220~\rm{GeV},~350~\rm{GeV},~500~\rm{GeV}$ and $\met>120~\rm{GeV},~220~\rm{GeV},~350~\rm{GeV},~500~\rm{GeV}$. We find that the SR1 give the most stringent constraint, and the result is shown in Fig.~\ref{fig:collider_q3}. The black curve is the cross section of $pp\to q_H\bar{q}_H j \to q \bar{q} A_H A_H j$ after imposing cuts in signal region of SR1. The grey shaded region represents the exclusion of mono-jet plus MET search.  We can see that the $f$ value below about $1.4$ TeV is excluded, which corresponds to exclusion of $m_{A_H} \lesssim 215~\rm{GeV}$.

\section{Conclusion}
\label{sec: con}
 
In this paper, we study the phenomenology of dark matter in the Littlest Higgs Model with T-parity, focusing on the coannhilation scenario and its implications at the LHC.  We find that, even though the T-odd partner of neutrino $\nu_H$ can be the lightest T-odd particle and satisfy the current measurement of dark matter relic abundance, it is ruled out by the direct dark matte search experiments because of its large cross section when scattering with nucleon. The T-odd heavy photon $A_H$ is therefore the only suitable dark matter in the model.    
For pair annihilation cross section of $A_H$ to explain the observed dark matter relic abundance with the Higgs boson being $125$ GeV, the mass of $A_H$ should be just slightly heavier than half of Higgs boson mass. However, the direct detection of dark matter of current data from LUX  disfavors this case. When $m_{A_H}$ becomes heavier, coannihilation contributions from T-odd fermions must be taken into account in order to fit the relic abundance measurement.    

For the case of coannihilation with light T-odd leptons, the existing LEP limit for heavy charged lepton search excludes the region where $f\lesssim 650~\rm{GeV}$, which corresponds to $m_{A_H}\lesssim 92~\rm{GeV}$. The current dark matter direct search result from LUX 2013 requires $m_{A_H}\gtrsim 110~\rm{GeV} $  while the future projected sensitivity of XENON 1T has potential to examine the whole parameter space. Since the T-odd leptons are as light as dark matter, the production cross section of T-odd lepton pairs at the LHC can be as  large as $O(10)~\rm{pb}$. However, the charged lepton in the final sate  is very soft since the mass difference between T-odd lepton and dark matter  is small. Therefore the direct searches at the LHC is very challenging and the current  searches for heavy charged particles have no constraints on these light T-odd leptons. For the scenario of T-odd quark coannihilation, we have to consider the situation that T-odd partner of top quark $t_H$ is much heavier than the T-odd partners of the first two generations of SM quarks and is irrelevant in all the results we have shown here. Otherwise, $t_H$ will be stable enough to generate displaced vertex signature at the LHC, which is inconsistent with data. When T-odd quarks are light, they contribute significantly to the $A_H$-nucleon elastic scattering cross section. The current dark matter direct search result from LUX 2013 imposes a constraint on $m_{A_H}\lesssim  70~\rm{GeV}$ while the projected result of LUX 2014 has the sensitivity for $m_{A_H} \lesssim 200~\rm{GeV}$. The future XENON 1T has  potential to testify the whole parameter space except for $m_{A_H}\sim 80~\rm{GeV}$ where a large destruction effect  by T-odd quarks happens.  Similar to the case of T-odd leptons, even though the LHC can produce the light T-odd quarks copiously, with the total production cross section as large as $10^{4}$ pb for $f\sim 500~\rm{GeV}$, the final state jets from T-odd quark decay are very soft, and therefore, it very difficult to directly search for these light T-odd quarks at the LHC. However, the production of  a T-odd quark pair plus one jet contributes significantly to mono-jet plus missing energy signature and the current data from ATLAS set a stringent constraint that disfavors $f\lesssim 1.4~\rm{TeV}$, corresponding to $m_{A_H}\lesssim 215~\rm{GeV}$.
 
 In summary, $A_H$ dark matter, the T-odd partner of photon,  in Littlest Higgs Model with T-parity fits the  measurement of dark matter relic abundance and can satisfy the direct search results well when coannihilaiton processes are considered. The current LHC mono-jet plus missing energy data sets a strong constraint on $A_H$ mass in the T-odd quark coannihilation scenario. For T-odd lepton coannihilaiton case, the LHC has no constraint while the direct detection of dark matter can be sensitive.  Combining the future data from dark matter direct search experiments and from LHC, we should be able to fully testify the whole interesting parameter of the model.   
 
\begin{acknowledgments}~The work of C.R.C. is supported in part by the National Science Council under Grant No.~NSC 102-2112-M-003-001-MY3. The work of M.C.L. is supported in part by the Grant No. 102-2815-C-003-034-M. H.C.T. acknowledges for the hospitality of ASIoP and NTNU.

\end{acknowledgments}

\end{document}